\documentclass[conference]{IEEEtran}

\usepackage[
backend=biber,
style=ieee,
sorting=ynt
]{biblatex}
\usepackage{amsmath,amssymb,amsfonts}

\usepackage{booktabs}
\usepackage{multirow}
\usepackage[normalem]{ulem}
\useunder{\uline}{\ul}{}
\usepackage{lscape}
\usepackage{longtable}
\usepackage{array}
\usepackage{tabularray}

\usepackage{algorithmic}
\usepackage{graphicx}
\usepackage{textcomp}
\usepackage{xcolor}

\usepackage[breaklinks]{hyperref}
\hypersetup{
    colorlinks=true,
    citecolor=blue,
    linkcolor=violet,
    urlcolor=teal,
    pdftitle={SmartphoneDemocracy: On-Device Privacy-Preserving E-Voting System with Decentralized Infrastructure and European Digital Identity Verification},
    pdfpagemode=FullScreen,
}

\def\BibTeX{{\rm B\kern-.05em{\sc i\kern-.025em b}\kern-.08em
    T\kern-.1667em\lower.7ex\hbox{E}\kern-.125emX}}
\begin{document}

\title{SmartphoneDemocracy: Privacy-Preserving E-Voting on Decentralized Infrastructure using Novel European Identity}

\author{\IEEEauthorblockN{Michał Jóźwik}
\IEEEauthorblockA{\textit{Delft University of Technology} \\
Delft, The Netherlands \\
M.G.Jozwik@student.tudelft.nl}
\and
\IEEEauthorblockN{Johan Pouwelse}
\IEEEauthorblockA{\textit{Delft University of Technology} \\
Delft, The Netherlands \\
J.A.Pouwelse@tudelft.nl}
}

\maketitle

\begin{abstract}
The digitization of democratic processes promises greater accessibility but presents challenges in terms of security, privacy, and verifiability. Existing electronic voting systems often rely on centralized architectures, creating single points of failure and forcing too much trust in authorities, which contradicts democratic principles. This research addresses the challenge of creating a secure, private e-voting system with minimized trust dependencies designed for the most versatile personal device: the smartphone. We introduce SmartphoneDemocracy, a novel e-voting protocol that combines three key technologies: the emerging European Digital Identity (EUDI) Wallet for Sybil-resistant identity verification, Zero-Knowledge Proofs for privacy-preserving validation, and a peer-to-peer blockchain (TrustChain) for a resilient, serverless public bulletin board. Our protocol enables voters to register and cast ballots anonymously and verifiably directly from their smartphones. We provide a detailed protocol design, a security analysis against a defined threat model, and a performance evaluation demonstrating that the computational and network overhead is feasible for medium- to large-scale elections. By developing and prototyping this system, we demonstrate a viable path to empower citizens with a trustworthy, accessible, and user-controlled digital voting experience.
\end{abstract}


\section{Introduction}
\label{sec:introduction}

The digitalization of democratic processes has immense potential for increasing accessibility and efficiency, but electronic voting (e-voting) systems face persistent and critical challenges. Ensuring security against tampering, guaranteeing voter privacy (anonymity and unlinkability), maintaining public verifiability, and preventing coercion remain significant hurdles \cite{evotingBlockchain2023}. Furthermore, many existing e-voting solutions rely on centralized infrastructure, creating single points of failure, potential censorship bottlenecks, and requiring voters to trust central authorities, thereby undermining the core tenets of democratic power distribution.

Current e-voting research often explores blockchain, but frequently within centralized or semi-centralized models \cite{electionblock2021}, which still depend on trusted intermediaries for vote casting, tallying, or identity verification. While enhancing transparency compared to traditional systems, these approaches may not scale effectively, can still leak metadata compromising anonymity, and often neglect the practicalities of user interaction, particularly on ubiquitous devices like smartphones. The challenge intensifies when considering global-scale participation, where centralized systems become logistically and computationally infeasible, and trust assumptions break down. A truly democratic digital voting system necessitates a shift towards fully decentralized architectures operated by the participants themselves.

Three key technological advancements offer a path forward. 
First, the European Union Digital Identity (EUDI) Wallet framework \cite{eidas2} represents a significant step toward standardized, user-controlled, and verifiable digital identity. It provides a mechanism for citizens to hold digitally signed attestations (Verifiable Credentials) about their identity and eligibility, issued by trusted authorities (Identity Providers), without relying on a central database during the voting act itself. 
Second, Zero-Knowledge Proofs \cite{goldwasser1989knowledge} provide the cryptographic means for a voter (prover) to convince a verifier (the system or other participants) that they are eligible to vote and have constructed their vote correctly, without revealing their identity or specific choice. Efficient ZKP schemes are becoming increasingly practical, even in resource-constrained environments such as smartphones. 
Third, peer-to-peer (P2P) networks, particularly blockchain-based systems designed for resource efficiency, provide a substrate for decentralized data storage, validation, and consensus, eliminating the need for central servers. 
A concise architecture, suitable for running on consumer devices, would align well with a smartphone-centric approach.

With SmartphoneDemocracy, we propose a new solution to the well-known topic of electronic voting, one that, for the first time, utilizes smartphones for the majority of its processing and integrates an emerging EUDI wallet for identity solutions.  
The primary contribution of this research is the development of a new state-of-the-art e-voting protocol. We redefine the current baseline for trust required in e-voting by utilizing modern cryptography and ideas of past designs, while lowering the trust requirements and the infrastructure required to organize elections without direct government interference.

\section{Problem description}
\label{sec:problem_description}

The goal of any electronic voting system should be to strengthen democracy by making it more accessible and transparent. However, most current approaches fail a fundamental test: they require citizens to place their trust in a central authority to manage the election. This reliance on central servers and institutions creates single points of failure and control, which are contrary to the democratic ideal of distributed power. Our research confronts this primary issue by asking: how can we conduct a secure and fair election without a central coordinator, relying instead on a system operated by the participants themselves?

This pursuit of a truly decentralized election immediately presents a critical challenge: establishing who is eligible to vote. In a traditional system, a government-managed voter list solves this problem. However, in a distributed system without a central authority, there is no definitive list. This creates a foundational vulnerability. How can the system ensure that every participant is a unique, eligible voter and prevent a single individual from casting multiple fraudulent votes? The problem is not just about having a strong, verifiable digital identity, but also about how to utilize that identity to prove eligibility within the network without requiring permission from a central server.

Furthermore, this decentralized framework must resolve the inherent conflict between voter privacy and public verifiability. For an election to be legitimate, two things must be true: each voter's choice must remain completely secret, and the final tally must be publicly auditable to prove the result is accurate. These two principles are naturally at odds. A fully transparent system could compromise ballot secrecy, while a perfectly secret system could hide fraudulent counting. The scientific challenge is therefore to design a method that allows each voter to prove their ballot was correctly cast and counted, without revealing any information about who they are or how they voted.

Finally, these solutions to identity and privacy must be practical for widespread use. The advanced cryptographic techniques required to achieve these goals can be computationally intensive. If the system is to be run on citizens' personal devices, it must be efficient enough to provide a smooth and accessible experience. A theoretically perfect system that is too slow, too complex, or drains too much battery ultimately fails in its goal of empowering voters. The problem, therefore, is to balance the need for robust security with the real-world performance constraints of technology that is available to everyone. Our work aims to deliver a unified solution that resolves these tensions, enabling a voting process that is secure and verifiable precisely because it does not depend on trust in any single institution.

\section{Background}
\label{sec:background}

To establish the technological foundation for SmartphoneDemocracy we need to examine five key areas: the current state of digital democracy and e-voting challenges, the emerging European Digital Identity infrastructure that enables Sybil-resistant authentication, Zero-Knowledge Proof systems that provide privacy-preserving validation, peer-to-peer networks that eliminate central points of control, and advanced cryptographic protocols that ensure ballot secrecy and system integrity. Together, these technologies create the building blocks for our smartphone-centric, decentralized voting protocol.

\subsection{Digital Democracy and E-Voting}
\label{sub:digital_democracy}
The concept of digital democracy has evolved significantly with the advancement of information and communication technologies. E-voting systems, in particular, have emerged as a promising avenue for enhancing democratic participation. However, the implementation of secure and reliable e-voting systems faces numerous challenges, even when in principle all elements of the democratic process could be improved (see Table \ref{tab:voting_diff}). Existing e-voting solutions often struggle to balance security, privacy, and verifiability, especially since the legal principles governing elections are applied with much stricter scrutiny to high-tech systems than to their traditional paper counterparts \cite{maurer2016voting, benabdallah2022analysis}. Furthermore, one might assume that the introduction of e-voting systems would increase democratic participation in society; however, case studies on existing systems show that the percentage of people attending elections remains constant in countries with good voting accessibility \cite{estonia_voting}.  
The case of Estonia's nationwide e-voting system, while demonstrating long-term adoption, also highlights persistent concerns from security experts regarding its centralized architecture and the potential for large-scale, undetectable fraud.
Studies show that while voters appreciate the convenience, trust in the security of electronic systems remains a significant barrier compared to traditional paper ballots \cite{alvarez2013voting, powell2012voting}.
The introduction of blockchain technologies has great potential to advance this field. However, it still needs to resolve its challenges and controversies regarding integrity, scalability, and transparency to be considered trustworthy for democratic elections \cite{evotingBlockchain2023}.

\begin{table*}
\centering
\caption{Comparison of traditional paper-based elections and a target e-voting system.}
\label{tab:voting_diff}
\begin{tblr}{
  width = \linewidth,
  colspec = {llQQQ[c]},
  row{1} = {c, m},
  cell{2}{1} = {r=2}{},
  cell{4}{1} = {r=2}{},
  vline{2-5} = {-}{},
  hline{2,4,6-8} = {-}{},
  hline{3,5} = {2-5}{},
}
\textbf{Stage}        & \textbf{Action/Property}  & \textbf{Traditional Elections}             & \textbf{Ideal E-voting}                              & \textbf{Upgrade?} \\
\textbf{Registration} & \textbf{Eligibility}      & Manual check of physical document          & Cryptographic proof from a digital wallet            & Yes                   \\
                      & \textbf{Verification}     & Human comparison against a list            & Automated verification on a public ledger            & Yes                   \\
\textbf{Voting}       & \textbf{Choice selection} & Pen and paper; possible to spoil ballot    & Constrained input; invalid votes prevented by design & Yes                   \\
                      & \textbf{Ballot casting}   & Physical ballot box; risk of observation   & Encrypted vote cast from a secure device             & Yes                   \\
\textbf{Tallying}     & \textbf{Counting}         & Manual, slow, prone to human error         & Automated, rapid, and deterministic counting         & Yes                   \\
\textbf{Verification} & \textbf{Integrity}        & Relies on trust in local committee members & Publicly verifiable via cryptographic primitives     & Yes                   \\
\textbf{Publication}  & \textbf{Results}          & Announced by a central official entity     & Publicly available on an immutable ledger            & Yes                   
\end{tblr}
\end{table*}

\subsection{EUDI Wallet and Decentralized Identity}
\label{sub:eudi_wallet}

A fundamental challenge in designing secure digital systems is preventing Sybil attacks, where a single adversary creates multiple fake identities to gain disproportionate influence \cite{douceur2002sybil}. Establishing a robust, unique, and user-controlled digital identity is therefore paramount. As society's reliance on the internet grows and privacy becomes a more pressing concern, research has shifted towards more decentralized solutions for identity management.

A prominent, regulation-driven initiative is the European Digital Identity Wallet, mandated by the EU's revised eIDAS 2.0 regulation \cite{eidas2}. The EUDI Wallet provides every EU citizen with a secure, user-controlled platform for their official digital identity, operating on the principles of Self-Sovereign Identity (SSI) \cite{preukschat2021ssi}. This paradigm grants individuals sole custody of their identity data, enabling them to present digitally signed Verifiable Credentials (VCs) to prove specific attributes (e.g., citizenship in a particular country) without disclosing personal information.

This government-backed approach can be contrasted with private-sector initiatives tackling the same Sybil problem. World (formerly known as WorldCoin), for example, establishes a globally unique identity network by using biometric iris scans to issue a "World ID," a method that has sparked significant debate regarding data privacy \cite{world2025}. Other projects, such as BrightID, use social verification, creating a "web of trust" where users vouch for one another's uniqueness \cite{brightid2022}.

\subsection{Zero-Knowledge Proofs}
\label{sub:zkp}
Zero-Knowledge Proofs are cryptographic protocols that allow one party (the prover) to prove to another party (the verifier) that a statement is true without revealing any information beyond the validity of the statement itself \cite{goldwasser1989knowledge}. ZKPs have numerous applications in cryptography, including privacy-preserving authentication and secure multi-party computation. When they were first introduced, they were primarily used as a theoretical concept, which later gained popularity with the introduction of blockchain technology and a greater emphasis on the privacy of digital systems. In e-voting, ZKPs are invaluable for proving statements like: "I know the secret that opens this registered commitment," "This ciphertext encrypts a valid choice from the official list," or "I correctly computed this partial decryption share," all without exposing the secret, the vote, or the decryption key.

In recent times, with the increased demand for ZKP-based solutions, further research has been conducted to discover more concise, efficient, and simpler schemes that possess similar security properties. 
Several families of ZKPs exist, each with different trade-offs, which include, but are not limited to:
\begin{itemize}
    \item zk-SNARKs (Zero-Knowledge Succinct Non-Interactive Argument of Knowledge) offer extremely small proof sizes and fast verification times, making them ideal for blockchain and mobile applications. Their primary drawback is the need for a one-time "trusted setup" for each program (circuit) being proven \cite{bitansky2012extractable}.
    \item zk-STARKs (Zero-Knowledge Scalable Transparent Argument of Knowledge) eliminate the need for a trusted setup, offering "transparent" and quantum-resistant security. However, they result in significantly larger proof sizes compared to SNARKs \cite{ben2018scalable}.
    \item Bulletproofs also require no trusted setup and produce very small proofs, but their proving and verification times are slower than SNARKs, especially for complex statements \cite{bunz2018bulletproofs}.
\end{itemize}
The choice of ZKP scheme depends heavily on the application's constraints, such as proof size, verification cost, and tolerance for a trusted setup.

\subsection{Peer-to-Peer Networks and Distributed Systems}
\label{sub:p2p}

Peer-to-peer (P2P) networks are distributed systems in which participants (peers) interact directly to share data and resources, thereby eliminating the need for a central server. This architecture is inherently resilient to single points of failure and resistant to censorship, as no single entity controls the network's operation. Fundamentally, P2P systems shift the basis of trust from a central intermediary to a transparent and verifiable protocol, often implemented as a blockchain or Distributed Ledger Technology (DLT). This has been most commonly used in systems such as Napster, BitTorrent, and Bitcoin.

This technological foundation enables novel forms of distributed governance, where rules are encoded into the system and decisions are made collectively by participants. A prominent example is the Decentralized Autonomous Organization (DAO), an entity governed by smart contracts and community consensus rather than a traditional management hierarchy \cite{buterin2013ethereum}. This stack of technologies provides the essential building blocks for applications that demand high integrity, transparency, and decentralized control. The Ethereum blockchain has explored the foundational concepts \cite{buterin2013ethereum}, which have since brought this governance model into the mainstream of blockchain technologies.

\subsection{Security and Anonymity in Cryptographic Protocols}
\label{sub:security_anonymity}

Modern digital infrastructure, from secure communication to e-commerce and digital identity, is fundamentally reliant on cryptographic protocols. These protocols provide the essential guarantees of confidentiality, integrity, and authenticity that underpin trust in the digital world. Beyond these basics, advanced cryptographic techniques enable complex, privacy-preserving interactions that were previously impossible.

A particularly powerful tool for privacy-preserving data analysis is Homomorphic Encryption (HE). HE schemes uniquely allow for mathematical operations, such as summation, to be performed directly on encrypted data (ciphertexts) \cite{paillier1999public}. This enables untrusted third parties, such as cloud servers or public ledgers, to process sensitive datasets without ever decrypting them, thereby preserving the privacy of the underlying information. For example, a system could calculate the total sum of financial transactions or the final tally of an election without ever exposing the individual values. This principle is often extended into a threshold cryptosystem, where the decryption key is split among multiple independent authorities. This ensures that no single entity can decrypt the final result, providing a robust defense against coercion and unilateral control.

For managing authorization and credentials anonymously, specialized signature schemes are employed. The BBS signature scheme, for instance, is explicitly designed for verifiable credentials \cite{looker2023bbsdraft, boneh2004short}. It allows an authority to sign a set of multiple attributes at once. Its primary feature is selective disclosure, where the holder can later generate a proof revealing only a chosen subset of these attributes while keeping the rest hidden. This proves possession of a valid credential without revealing the user's whole identity. It also has a highly relevant property of aggregation, which allows joining multiple signatures together and verifying them all at once, resulting in significant performance boosts in large-scale verifications.

As a more comprehensive alternative, Secure Multi-Party Computation (SMPC) enables groups to jointly compute any function over their private inputs. While extremely powerful, SMPC often incurs high communication overhead, which can present scalability challenges in large-scale, decentralized environments.

\section{Proposed Framework}
\label{sec:framework}

The transition from centralized to decentralized voting is not merely a technological upgrade; it represents a fundamental reimagining of democratic participation itself. Rather than just digitizing existing institutional processes, we embrace a new premise: that citizens should directly control the infrastructure of their own elections. This framework orchestrates a carefully designed set of cryptographic elements, where voters bootstrap their own eligibility through digital credentials, anonymously participate using privacy-preserving proofs, and collectively maintain the integrity of the results without surrendering control to any single authority. The elegance lies not in the individual cryptographic primitives, which are well-established, but in their synthesis into a system that runs entirely on the citizens' mobile devices, transforming every smartphone into both a private voting booth and a node in the democratic infrastructure. What emerges is a protocol that treats verifiability as a public good and privacy as an uncompromisable right, while remaining practical enough for deployment at a national scale.

\subsection{Election Phases}

The protocol proposed in this research is structured into four distinct, sequential phases. This phased approach is not arbitrary; it is a deliberate cryptographic parallel to the fundamental stages of any legitimate democratic election. Each phase serves a crucial function, ensuring that the digital process inherits and enhances the security, fairness, and verifiability of its physical counterpart.

\paragraph{Election Setup} This initial phase serves as the foundational administrative step, analogous to an electoral commission formally announcing an election, defining its rules, and printing the official ballots. In our digital protocol, this translates to establishing the cryptographic ground truth for the entire process. The election parameters, public keys for encryption, and verification keys for proofs are published on the public ledger. This step is essential because it creates a binding public contract that all participants can trust and rely upon. It prevents any single party from changing the rules or cryptographic keys during the election, ensuring a fair and universally agreed-upon foundation before any votes are cast.

\paragraph{Voter Registration} The registration phase directly addresses the "one person, one vote" principle. In a traditional election, a citizen presents identification at a polling station, an official verifies their eligibility, and their name is crossed off a list to prevent them from voting again. Our protocol's registration phase is the cryptographic equivalent of this check-in process. It is designed to solve the core privacy paradox: how to verify a voter's eligibility without linking their real-world identity to their eventual ballot. By using the EUDI credential off-chain to obtain a single-use, anonymous on-chain credential (via the BBS signature), this phase acts as a cryptographic "air gap", providing proof of eligibility while breaking the chain of evidence that could compromise voter anonymity. This design introduces a centralization point through the trusted Verifier, a limitation we acknowledge as a temporary practical compromise. While a fully decentralized protocol would distribute this verification responsibility (as outlined in Section \ref{sub:future_work}), our current approach strikes a balance between immediate implementability and core security requirements.

\paragraph{Voting} Once a voter is registered and possesses an anonymous right to vote, the voting phase facilitates the secret expression of their choice. This is the digital equivalent of entering a private polling booth, marking a ballot, and placing it in a sealed ballot box. Our protocol achieves this through a combination of cryptographic tools. Homomorphic Encryption acts as the "sealed envelope," ensuring the vote's content remains secret even when published on a public ledger. The vote nullifier acts as the "ballot stub", a unique, publicly visible proof that the voter has used their right to vote, making double-voting impossible. Finally, the Zero-Knowledge Proof serves as an affidavit, guaranteeing that the encrypted vote is validly formed and cast by a legitimately registered participant, all without revealing any secret information.

\paragraph{Tallying} The final phase mirrors the public counting of ballots under the watchful eye of observers. Instead of relying on trust in human counters, our protocol relies on the verifiable mathematics of homomorphic encryption and threshold cryptography. Anyone can independently perform the homomorphic summation of all encrypted votes, and the threshold decryption mechanism distributes the power to reveal the final result, preventing any single entity from manipulating or censoring the outcome. This phase transforms the trusted, manual process of counting into a trustless, automated, and publicly verifiable computation, providing the guarantee of the election's integrity.

\subsection{System Architecture}
\label{sub:system_architecture}
The SmartphoneDemocracy framework is a layered, P2P system where each component serves a distinct purpose in achieving our goals of security, privacy, and decentralization. At its base is a lightweight P2P network that serves as a public bulletin board. Additionally, a cryptographic protocol coordinates the actions of voters, who use their smartphones to interact with the system. Voter eligibility is bootstrapped through the external EUDI Wallet ecosystem. The interaction between these components is designed to minimize trust and maximize public verifiability.

For the blockchain, we will be using TrustChain \cite{otte2020trustchain} technology as the primary provider of tamper-proof and verifiable data structure. It is a blockchain solution developed at TU Delft that enables the creation of trusted transactions among strangers without requiring central control. Its architecture avoids a global consensus on a total ordering of transactions, making it efficient for the kind of append-only data necessary for an election. Each voter's actions (registration, voting) are recorded as transactions on their personal chain, which are then gossiped and cross-validated throughout the network. This provides a tamper-evident and universally accessible record of the election proceedings, eliminating the need for a central server.

\subsection{Assumptions}
In this paper, we focus on binary (yes/no) majority voting. The protocol can be extended to other voting schemes, but we limit our scope to this common form. The final interpretation of results (e.g., handling ties) is left to the election organizers. It is not enforced by the protocol, with an option to provide further details in the election configuration.

Our system operates under the key assumptions, which are listed in the Table \ref{tab:assumptions}.

\begin{table}[ht]
\centering
\caption{Assumptions of the SmartphoneDemocracy system.}
\label{tab:assumptions}
\begin{tblr}{
  width = \linewidth,
  colspec = {cX},
  column{1} = {m},
  vline{2} = {-}{},
  hline{2-8} = {-}{},
}
\textbf{ID} & \textbf{Description}                                                                                                                                                                                                           \\
\textbf{A1} & Trusted Identity Infrastructure: Official Identity Providers issue non-forgeable, EUDI-compatible Verifiable Credentials for eligibility only to legitimate voters. The EUDI Wallet application on the user's phone is secure. \\
\textbf{A2} & One Person, One Vote: Each eligible voter can successfully complete the registration protocol exactly once per election to obtain their anonymous credentials.                                                                 \\
\textbf{A3} & Open Proposal Creation: Any registered participant is permitted to create a new voting proposal. (This can be restricted by policy in a real deployment.)                                                                      \\
\textbf{A4} & Simple Voting Schemes: Proposals involve binary selection where the winner is determined by plurality. (Could be expanded, but outside of scope for this paper)                                                                \\
\textbf{A5} & Capable User Devices: Voters possess smartphones capable of running the EUDI Wallet, the voting application, and performing ZKP computations within an acceptable timeframe. The device's OS and hardware are not compromised. \\
\textbf{A6} & P2P Network Liveness: The underlying P2P network (TrustChain/IPv8) is operational and accessible, with sufficient honest participation to ensure data propagation and persistence as per its security model.                   \\
\textbf{A7} & Secure Cryptography: All underlying cryptographic primitives (ZKPs, HE, digital signatures, hash functions) are computationally secure against the modeled adversary.                                                          
\end{tblr}
\end{table}

\subsection{Threat Model}
To analyze the security of our protocol, we define the capabilities of our adversary. We consider a powerful but computationally bounded adversary who can control a fraction of the participants and observe all network traffic. The primary adversarial roles are:
\begin{itemize}
    \item Malicious Voter: An eligible voter who attempts to break the protocol rules, for example, by trying to vote more than once, deanonymize other voters, or disrupt the tallying process.
    \item External Eavesdropper: An entity that monitors all network communications to link voter identities to their pseudonymous actions (registration, voting) through traffic analysis.
    \item The Coercer: An adversary who attempts to force a voter to vote for a specific candidate, abstain from voting, or reveal how they voted. The coercer may have access to the voter's device after the fact.
    \item Compromised P2P Peers: A coalition of malicious nodes in the P2P network that may attempt to censor valid transactions or refuse to propagate data. We assume they do not constitute a majority sufficient to break the underlying security of TrustChain.
\end{itemize}
We assume the EUDI Identity Provider and the voter's smartphone itself are trusted components, as per assumptions \textbf{A1} and \textbf{A5} from Table \ref{tab:assumptions}. Compromise of these components is considered an out-of-scope, systemic failure. Our goal is to design a protocol that remains secure even if all other components (the network, other voters) are adversarial.

\subsection{Stakeholders}

The proposed system involves three main groups of stakeholders:
\begin{itemize}
    \item Voters: Citizens who use the smartphone application to register their eligibility and cast their ballots. Their primary interest is in a system that is usable, private, and trustworthy.
    \item Identity Providers (and EUDI Infrastructure): The government agencies or other trusted entities responsible for issuing the Verifiable Credentials that prove a voter's eligibility. They are stakeholders in the secure and interoperable use of their credentials.
    \item Public Verifiers (Any Participant): Any individual or organization, including voters, auditors, or journalists, who independently downloads the public data from TrustChain to verify the integrity of the election process and its outcome. They play a crucial role in ensuring transparency and building trust in the system.
\end{itemize}

\section{Protocol Design}
\label{sec:protocol_design}

The SmartphoneDemocracy protocol is divided into four main phases, as described in Section \ref{sec:framework}. It orchestrates several cryptographic primitives to meet the requirements of privacy, security, and verifiability. The core components include BBS signatures for issuing privacy-preserving voting credentials, which possess native zero-knowledge properties; Zero-Knowledge Proofs for general-purpose validation; additively Homomorphic Encryption for ballot secrecy; digital signatures for authenticity; and a P2P ledger (TrustChain) as a public bulletin board.

\subsection{Choice of Cryptographic Primitives}
The selection of cryptographic primitives is a deliberate engineering decision, balancing demands of security and privacy against the practical constraints of a smartphone-centric, peer-to-peer environment. The primary goals guiding our choices were minimizing on-chain data size, ensuring fast verification by network peers, and maintaining a low computational burden on the voter's device.

\paragraph{BBS Signatures for Credential Issuance}
For the voter registration phase, we specifically chose the BBS signature scheme \cite{looker2023bbsdraft} over a general-purpose ZKP. The task of issuing a credential and later proving possession of it with selective disclosure is the native function of BBS. This choice offers several advantages, such as: generating a BBS proof is computationally trivial, typically taking only milliseconds on a mobile device. This is orders of magnitude faster than generating a general-purpose ZKP, ensuring a seamless user experience during registration. BBS is a well-established standard in the decentralized identity community (e.g., W3C VC Data Model \cite{bernstein2024bbsw3c}). Adopting it aligns our protocol with a broader, well-vetted ecosystem, rather than relying on a custom, ad-hoc ZKP circuit for this common task. Finally, it eliminates the need to design, implement, and audit a complex ZKP circuit for proving credentials, reducing the attack surface and development overhead. The underlying choice of the \texttt{BLS12-381} \cite{barreto2002constructing} pairing-friendly elliptic curve can also be justified as it is efficient for digital signatures, as well as the zk-SNARK proof generation  \cite{groth2016size}, which is a crucial protocol element as well.

\paragraph{Threshold Homomorphic Encryption for Tallying}
To ensure ballot secrecy while allowing for a public, verifiable tally, we employ an additively homomorphic encryption scheme. Specifically, we select a threshold generalized Paillier cryptosystem, proposed by Damg{\aa}rd et al. in \cite{damgaard2010generalization}. Paillier is a well-analyzed and efficient scheme for addition, which is the main requirement for our tallying system. Furthermore, in their research, Damg{\aa}rd et al. describe the exact use case in electronic voting, the method for creating a ZKP necessary to prove encryption correctness, and a trusted dealer threshold scheme, all of which are relevant to this research.

There exist other generally powerful homomorphic schemes, which allow for fully homomorphic encryption (FHE), unlocking access to any efficiently calculable function \cite{gentry2009fully}. However, even modern FHE schemes (such as TFHE \cite{chillotti2020tfhe}) suffer from a significant ciphertext expansion factor, which renders them unusable in our storage-constrained scenario.  

The threshold property is essential for our decentralized architecture, as it distributes the decryption key among participants in shares. This removes the single point of trust and failure associated with a central tallying authority holding a master decryption key, which is a critical requirement for a permissionless system. The ideal setup involves a Distributed Key Generation (DKG) protocol (such as one proposed by Veugen et al. \cite{veugen2019implementation} or using suggested scheme from Damg{\aa}rd et al. other paper \cite{damgaard2001practical}), so no trusted dealer is needed even for the key creation. 

\paragraph{Groth16 zk-SNARK for Voting Proofs}
While BBS is ideal for registration, the voting proof (\(\pi_{\text{vote}}\)) involves more complex logic: it must link a commitment, a nullifier, and an encrypted vote into a single, valid statement. This requires a general-purpose ZKP system. We chose the Groth16 zk-SNARK protocol \cite{groth2016size} after analyzing the trade-offs against alternatives like zk-STARKs or Bulletproofs, taking into account existing research \cite{oude2024systematic}. The primary reason is that it generates extremely small proofs (under 200 bytes), a crucial advantage for on-chain applications where storage and bandwidth are limited. In a P2P network, this directly impacts scalability.
Verification of Groth16 proofs is exceptionally fast, which is crucial for network peers that must validate numerous incoming vote transactions. The primary drawback of Groth16 is its requirement for a trusted setup for each circuit. We acknowledge this limitation but consider it a necessary and acceptable trade-off to achieve the performance required for a large-scale mobile system. This trusted setup limitation can be mitigated through large-scale MPC ceremonies (e.g., Zcash's Powers of Tau \cite{bowe2017scalable}). Alternatives like zk-STARKs, while not requiring a trusted setup, produce proofs that are orders of magnitude larger, making them impractical for this specific architecture.

\subsection{Election Setup}
An authorized entity (e.g., an election organizer or any user per \textbf{A3}) initializes the election by performing the following steps:
\begin{itemize}
    \item Defines the election parameters: A unique election identifier $id_E$, the proposal text, and the set of valid vote choices $V_C$.
    \item Configures the OpenID4VP Verifier service, specifying the required VC type for eligibility and the public keys of the trusted issuers $PK_{Iss}$. \\
    \item Generates cryptographic keys:
        \begin{itemize}
            \item An additively Homomorphic Encryption (HE) key pair $(sk_{HE}, pk_{HE})$. For threshold decryption, the secret key $sk_{HE}$ is split into $N$ shares $\{sk_{HE, i}\}_{i=1}^N$ using a $(t, N)$ threshold scheme, where $t$ is the decryption threshold. A Distributed Key Generation (DKG) protocol is strongly recommended for generating these keys without relying on a single trusted dealer.
            \item ZKP parameters for the chosen zk-SNARK scheme, including the Common Reference String (CRS) and the verification keys $(vk_{vote}, vk_{share})$ for the voting and tallying phases, respectively.
        \end{itemize}
    \item Publishes the election configuration $Config_E$ to TrustChain. This transaction includes $id_E$, election rules, $pk_{HE}$, the threshold $t$, and the ZKP verification keys.
    \item Establishes a mechanism for the distribution of HE secret key shares $sk_{HE, i}$ to be claimed by registered voters during the registration phase.
\end{itemize}

\begin{figure*}[!ht]
    \centering
    \includegraphics[width=\textwidth]{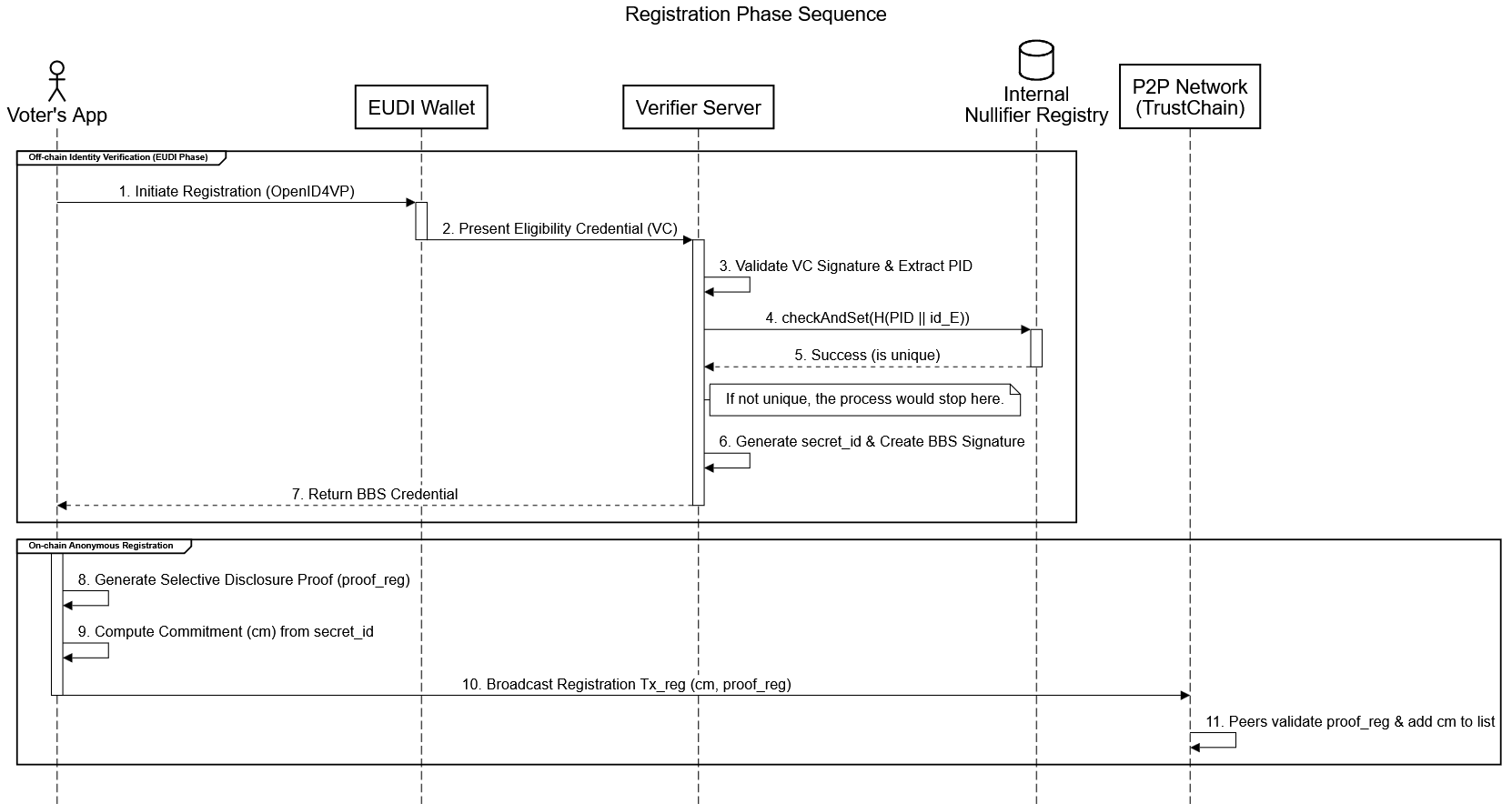}
    \caption{Registration sequence diagram illustrating the interaction between the Voter's App, EUDI Wallet, and the Verifier, culminating in a transaction to the P2P network.}
    \label{fig:registration_sequence}
\end{figure*}

\subsection{Voter Registration}
This phase enables a voter to use their official identity to gain an anonymous right to vote. It leverages BBS signatures to issue a cryptographic credential that can be utilized on-chain without revealing the voter's identity or linking their registration action across different services. For the primary protocol, we assume a single, trusted Verifier entity that maintains an internal nullifier registry to prevent duplicate credential issuance.

\begin{enumerate}
    \item Identity Verification (Off-chain): The voter's app initiates an OpenID4VP flow. The voter uses their EUDI Wallet to present their eligibility VC to the election's Verifier service. The Verifier validates the VC's signature against the trusted issuers list $PK_{Iss}$. 
    \item Internal Nullifier Check (Verifier): The Verifier extracts a unique Person Identifier (PID) from the VC and computes a verifier-side nullifier \(v_{nf} = H(\text{PID} \Vert \text{id}_E)\). It checks its internal database to ensure this \(v_{nf}\) has not been used before. If it has, the process stops. If not, it records \(v_{nf}\) and proceeds.
    \item BBS Credential Issuance (Verifier): The Verifier generates a new, high-entropy secret for the voter, \texttt{secret\_id}, and issues a BBS signature over a set of attributes:
    \(\{\text{secret\_id}, \text{election\_id}, \text{issuance\_timestamp}\}\).
    This signature, along with the original attributes, is sent to the voter's app.
    \item Proof Generation (Voter App): The voter's app uses the received BBS credential to generate a selective disclosure proof, denoted \(\text{proof}_{reg}\). This proof reveals the \texttt{election\_id} attribute publicly, while keeping the \texttt{secret\_id} and \texttt{issuance\_timestamp} attributes hidden. The proof cryptographically guarantees that the hidden attributes were part of the original set signed by the Verifier.
    \item On-chain Identifier Derivation (Voter App): The app uses the hidden \texttt{secret\_id} to create its on-chain anonymous identifiers:
        \begin{itemize}
            \item A cryptographic commitment \\ \(cm = \text{Commit}(\text{secret\_id})\).
            \item A vote nullifier \(nf_{\text{vote}} = H(\text{secret\_id} \Vert \text{id}_E)\). This will be used later during the voting phase to prevent double-voting.
        \end{itemize}
    \item TrustChain Submission (Voter App): The voter broadcasts a registration transaction \(Tx_{reg} = (cm, \text{proof}_{reg})\).
    \item Validation (TrustChain Peers): Any peer receiving this transaction verifies \(\text{proof}_{reg}\) against the Verifier's public BBS key and the public \texttt{election\_id}. If the proof is valid, they add the commitment \(cm\) to the list of eligible voters, $\mathbb{L}_{Commit}$.
\end{enumerate}

This process is the main novelty of the EUDI Wallet introduction, and the sequence diagram describing the registration is shown in Fig. \ref{fig:registration_sequence}.

\subsection{Voting}
Once registered with a valid commitment \(cm\) on the ledger, the voter can cast their encrypted vote anonymously. This phase leverages a general-purpose ZKP to prove the validity of the vote without linking it to the registration transaction.

\begin{enumerate}
    \item Vote Preparation (Voter App): The voter selects their choice \(v \in V_C\). The app retrieves the \texttt{secret\_id} corresponding to their registered commitment \(cm\).
    \item Vote Encryption (Voter App): The app encrypts the vote using the public homomorphic key: \(c = \text{Enc}_{pk_{HE}}(v)\).
    \item Vote Nullifier Calculation (Voter App): The app computes the vote nullifier \(nf_{\text{vote}} = H(\text{secret\_id} || id_E)\). This value is unique to the voter for this specific election and can only be generated by someone knowing the \texttt{secret\_id}.
    \item Voting Proof Generation (Voter App): The app generates a ZKP, \(\pi_{\text{vote}}\), using a general-purpose system, such as Groth16. This proof attests to a set of statements without revealing the underlying secrets. The prover demonstrates knowledge of a witness \(w_{\text{vote}} = (\text{secret\_id}, v)\) such that:
        \begin{itemize}
            \item The commitment \(cm = \text{Commit}(\text{secret\_id})\) exists in the public list of eligible commitments, \(\mathbb{L}_{\text{Commit}}\). (This can be proven efficiently using a Merkle proof against the root of \(\mathbb{L}_{\text{Commit}}\)).
            \item The vote nullifier \(nf_{\text{vote}}\) was correctly derived from this \texttt{secret\_id}.
            \item The ciphertext \(c\) is a valid HE encryption (as described by Damg{\aa}rd et al. \cite{damgaard2010generalization}) of a choice \(v\) from the allowed set \(V_C\).
        \end{itemize}
    \item TrustChain Submission (Voter App): The voter broadcasts a vote transaction \(Tx_{\text{vote}} = (id_E, c, nf_{\text{vote}}, \pi_{\text{vote}})\).
    \item Validation (TrustChain Peers): Peers verify the ZKP \(\pi_{\text{vote}}\) using the election's verification key \(vk_{\text{vote}}\). They also check that \(nf_{\text{vote}}\) has not already been published in the list of used vote nullifiers, \(\mathbb{L}_{\text{VoteNull}}\). This prevents double-voting.
    \item State Update (TrustChain): If valid, peers add the ciphertext \(c\) to the encrypted ballot box \(\mathbb{L}_{\text{BallotBox}}\) and add \(nf_{\text{vote}}\) to \(\mathbb{L}_{\text{VoteNull}}\).
\end{enumerate}

\subsection{Tallying}
After the voting period ends, the final result is computed through a publicly verifiable, two-step process.
\begin{enumerate}
    \item Homomorphic Summation (Public): Anyone can compute the aggregate ciphertext $C_{sum}$ by applying the homomorphic addition operation to all ciphertexts in the ballot box: $C_{sum} = \bigoplus_{c_i \in \mathbb{L}_{BallotBox}} c_i$. Due to the homomorphic property, $C_{sum}$ is an encryption of the sum of all plaintext votes.
    \item Threshold Decryption (Participants): To decrypt $C_{sum}$, a threshold $t$ of the original HE secret key shares are required. The participants who hold these shares (e.g., the first $N$ registered voters, or a designated committee) perform the following:
        \begin{itemize}
            \item Each participant $i$ computes a partial decryption share $\sigma_i = \text{PDec}_{sk_{HE, i}}(C_{sum})$.
            \item Each participant generates a ZKP $\pi_{share, i}$ proving they correctly computed their share $\sigma_i$ using the secret key $sk_{HE, i}$ corresponding to their role.
            \item Each participant publishes their partial decryption and its proof: $Tx_{share} = (id_E, \sigma_i, \pi_{share, i})$.
        \end{itemize}
    \item Final Result Combination (Public): Anyone can collect at least $t$ valid partial decryptions $(\sigma_i, \pi_{share, i})$ from the ledger, verify their proofs, and combine them to reveal the final plaintext tally $T$.
\end{enumerate}

\subsection{Public Verification}
Universal verifiability is a key outcome. Any interested party can download the entire election record from TrustChain and independently:
\begin{itemize}
    \item Verify all ZKPs ($\pi_{vote}, \pi_{share}$) and \(\text{proof}_{reg}\) to ensure every step was performed correctly.
    \item Check the uniqueness of nullifier ($nf_{vote}$) to confirm no double-voting occurred.
    \item Re-compute the homomorphic sum $C_{sum}$ from the public ballots.
    \item Re-compute the final tally $T$ from the published partial decryption shares.
\end{itemize}
This ensures the election's outcome is a direct and accurate consequence of the published votes, without relying on any single party.

\subsection{Implementation Details}
\label{sub:implementation_details}

\begin{figure*}[ht]
    \centering
    \includegraphics[width=0.3\textwidth]{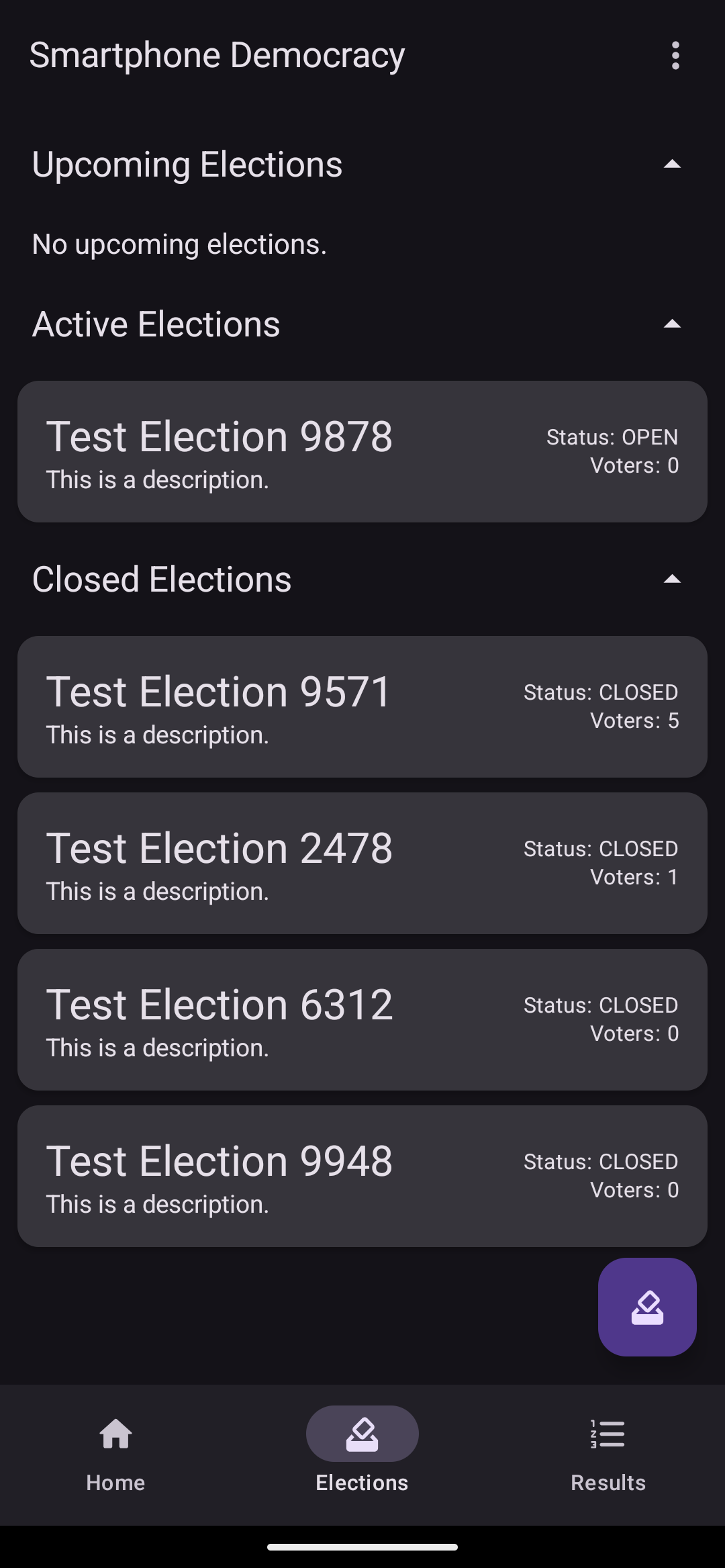}
    \includegraphics[width=0.3\textwidth]{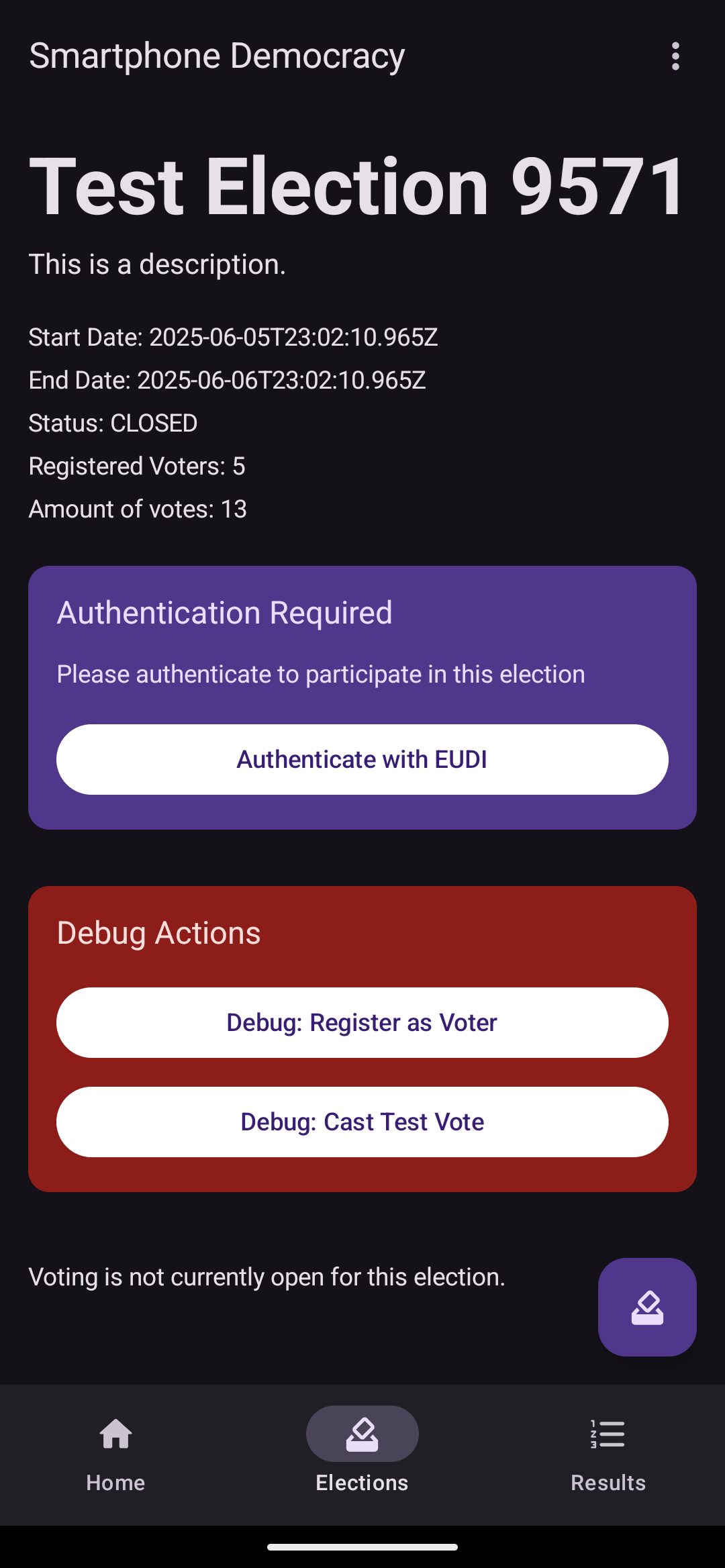}
    \includegraphics[width=0.3\textwidth]{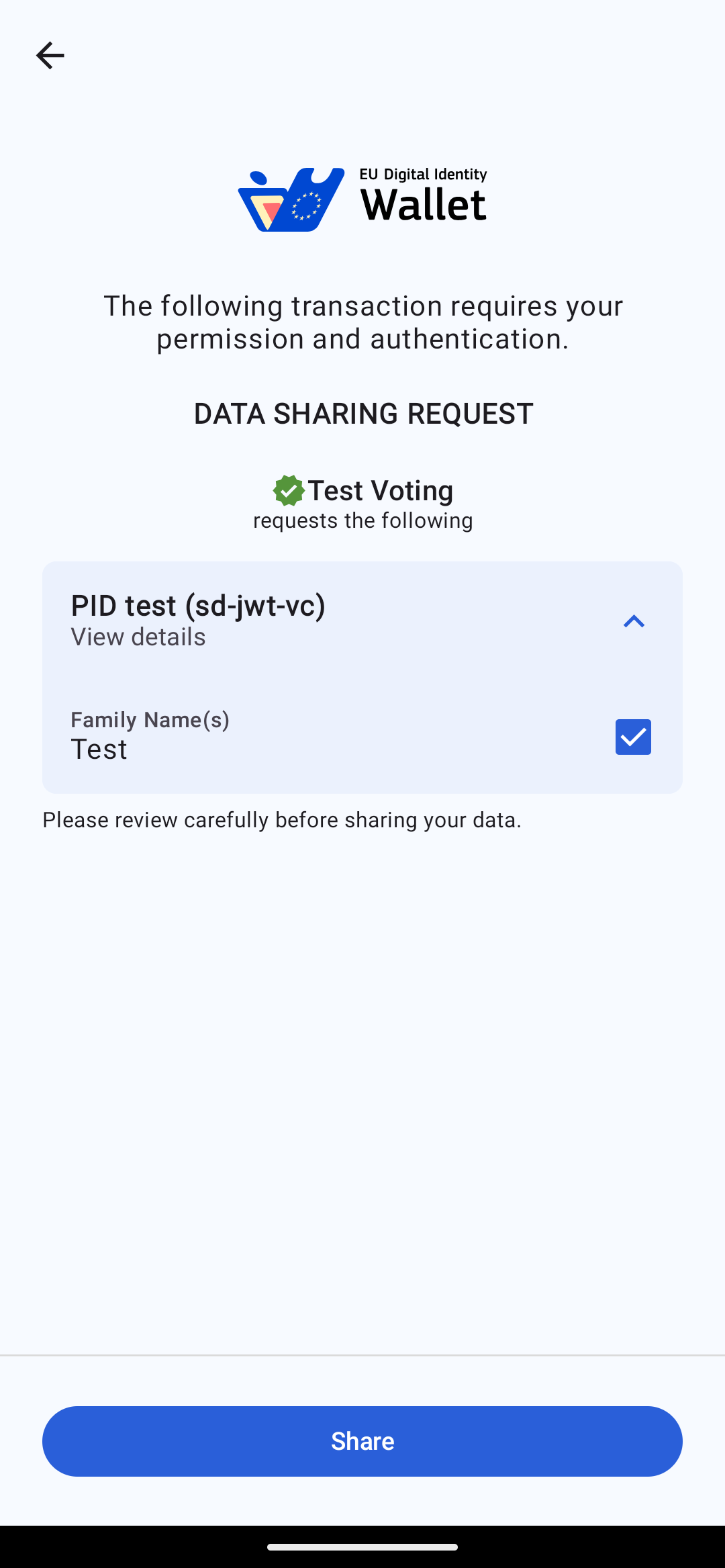}
    \caption{Example app screenshots displaying election list, election details, and EUDI Wallet confirmation, respectively.}
    \label{fig:app_usage}
\end{figure*}

To validate the practical feasibility of the SmartphoneDemocracy protocol, we developed a proof-of-concept prototype as an Android application. The choice of platform was guided by the generally open Android ecosystem and the availability of the Trustchain Super App\footnote{Available at: \url{https://github.com/Tribler/trustchain-superapp}.} project, which provides our foundational architecture interface. The application is written in Kotlin, the official language for modern Android development, and is required by the underlying \texttt{kotlin-ipv8} library that powers the peer-to-peer communication. To ensure a maintainable and performant architecture, we adopted a modern Android stack, utilizing Jetpack Compose for the user interface, Hilt for dependency management, and the Room persistence library for on-device caching of blockchain data. Fig. \ref{fig:app_usage} shows the example screenshots of the frontend implementation based on the described architecture.

For the identity verification, our architecture is designed to interface with the European Digital Identity ecosystem. The prototype assumes the user possesses a standard-compliant EUDI Wallet application on their device. The registration flow from our application initiates a request to this wallet using the OpenID4VP protocol. This triggers the user to present their eligibility credential to our server-side EUDI Verifier component. The core modification in our implementation lies in extending the standard Verifier logic. After successfully validating the credential, our Verifier performs the internal nullifier check and then issues the required BBS signature back to the application. This entire process is anchored in the trust placed upon the broader certification ecosystem, specifically the official Identity Providers responsible for issuing the initial eligibility credentials stored within the citizen's EUDI Wallet. All the code is based on the reference implementations available on the official project's GitHub organization: \url{https://github.com/eu-digital-identity-wallet/}.

A key architectural decision was to delegate all intensive cryptographic operations to a separate, high-performance native core, which interfaces with the Kotlin application via the Java Native Interface (JNI). Rust was the clear choice for this native core, selected for its strong memory safety guarantees without a garbage collector, its performance rivaling that of C++, and, critically, its vibrant and expanding ecosystem of cryptographic libraries. Our protocol design directly leverages this ecosystem, as the Kotlin (and overall Java) ecosystem lacks the required elements. For instance, the credential issuance is handled by a Rust implementation of the BBS signature standard \cite{looker2023bbsdraft} (with an alternative implementation with improvements and relation to other elements \cite{tessaro2023revisitingbbs, githubDockNetworkLibrary}). At the same time, the secure tallying is designed to utilize a library that provides the Paillier cryptosystem for homomorphic encryption. For the complex vote-casting proof, our design incorporates the powerful \texttt{arkworks} framework \cite{arkworks}, which implements the majority of underlying cryptographic primitives used in the ecosystem, with circuits defined in the accessible Circom language \cite{belles2022circom} and proof generation handled by a backend like \texttt{legogroth16} \cite{campanelli2019legogroth,githubDockNetworkLibrary}. This would also be integrated with proof aggregation tools like \texttt{snarkpack} \cite{gailly2021snarkpack}, which could significantly reduce on-chain verification costs in large-scale deployments, further enhancing the system's scalability.

Given the considerable engineering complexity involved, the proof-of-concept focused on implementing the most critical and novel components of the user-facing flow. The core Android application, the TrustChain peer-to-peer networking layer, and the registration process using the EUDI Wallet were partially implemented and tested. The complete integration of the homomorphic tallying and the verification of ZKP, however, remains a detailed design proposal. This was due to the significant challenges in securely managing the state across the JNI boundary within the available timeframe, as well as a lack of direct experience in Rust development. Nevertheless, this foundational work validates our architecture and opens clear avenues for future development, based on the proposed framework.

The code for the prototype is available on GitHub: \url{https://github.com/Eragoneq/trustchain-superapp/tree/smartphone-democracy}.

\section{Security Analysis}
\label{sec:analysis_evaluation}

This section examines how the protocol design achieves its core security and privacy objectives in the face of the adversary defined in our threat model. A more detailed risk analysis is available in Appendix \ref{app:proofs}.

\subsection{Eligibility and Uniqueness (Sybil Resistance)}
Uniqueness, the core defense against Sybil attacks, is enforced through a robust two-stage process that combines off-chain trusted verification with on-chain cryptographic checks.

First, eligibility and registration uniqueness are enforced off-chain by the trusted Verifier. Before issuing a credential, the Verifier checks an internal, private nullifier registry tied to the voter's unique Person Identifier (PID) from their EUDI credential. This ensures that each unique citizen can receive only one valid BBS voting credential for any given election, effectively preventing an individual from registering multiple times at the source.

Second, the one-vote-per-registrant rule (as per \textbf{A2}) is enforced on-chain through a cryptographic nullifier scheme. During registration, each voter is issued a unique \texttt{secret\_id} as a hidden attribute in their BBS credential. The vote nullifier, \(nf_{\text{vote}} = H(\texttt{secret\_id} || id_E)\), is derived from this secret. Since the \texttt{secret\_id} is inextricably linked to a single valid registration (represented by the commitment \(cm\)), any attempt to vote more than once would require reusing the same secret, thereby generating the exact same \(nf_{\text{vote}}\). This duplicate nullifier would be immediately identified and rejected by network peers checking against the public list of spent nullifiers, \(\mathbb{L}_{\text{VoteNull}}\). This cryptographically guarantees that each valid registration can result in exactly one valid vote.

\subsection{Anonymity and Unlinkability}
Anonymity is achieved by cryptographically severing the link between the voter's off-chain identity verification and their on-chain actions. The BBS signature scheme primarily accomplishes this.
\begin{itemize}
    \item During registration, the Verifier issues a BBS credential containing a high-entropy \texttt{secret\_id}.
    \item The voter generates a selective disclosure proof (\texttt{proof\_reg}) which proves they received a valid signature from the Verifier for the correct election. Still, crucially, it does not reveal the \texttt{secret\_id}.
    \item The voter's on-chain presence is defined only by the commitment \(cm = \text{Commit}(\text{secret\_id})\).
    \item Later, the vote transaction is linked to this registration only through knowledge of the \texttt{secret\_id}, which is proven inside the ZKP \(\pi_{\text{vote}}\).
\end{itemize}
An adversary observing the blockchain sees a registration proof and a later vote proof, but cannot link them without breaking the zero-knowledge property of the BBS scheme or the ZKP, or breaking the hiding property of the commitment scheme. The underlying \texttt{secret\_id} is never revealed.

Network-level anonymity remains vulnerable to timing correlation attacks. An adversary observing network patterns could potentially link registration and voting transactions from the same IP address. Mitigation strategies could include randomized transaction delays, Tor integration for network anonymity, or mixing services at the application layer. However, these additions would significantly complicate the user experience and mobile deployment.

\subsection{Ballot Secrecy}
The content of each vote $v$ is protected by the additively homomorphic encryption scheme. The vote is encrypted into a ciphertext $c$ on the voter's device before being transmitted. It remains encrypted on the public ledger. Only the final sum $C_{sum}$ is decrypted, and this requires a threshold $t$ of participants to cooperate. As long as fewer than $t$ tallying participants collude, no individual vote can be decrypted. The ZKP $\pi_{vote}$ ensures that $c$ encrypts a valid choice without revealing which one.

\subsection{Coercion Resistance and Receipt-Freeness}
Perfect coercion resistance remains an open challenge. Our protocol provides a significant degree of resistance through receipt-freeness. A voter cannot easily construct a "receipt" to prove to a coercer how they voted. The vote $c$ is just an encrypted ciphertext. The ZKP $\pi_{vote}$ is zero-knowledge, so it reveals nothing about the vote's content. While a coercer could observe the voter's screen, the cryptographic protocol itself does not produce an artifact that can be used as proof of a specific vote. Future work should consider employing more advanced schemes, such as those proposed by zkVoting \cite{cryptoeprint:2024/1003}.

\subsection{Universal Verifiability}
As described in Section \ref{sec:protocol_design}, the entire electoral process is publicly auditable. The presence of all transactions and proofs on the immutable TrustChain ledger allows any third party to perform a complete, independent verification of the election's integrity from start to finish.

\section{Performance and Scalability Evaluation}
\label{sub:performance_evaluation}

This section transitions from cryptographic theory to real feasibility, examining the protocol's computational performance, network requirements, and usability to validate its readiness for large-scale deployment.

\subsection{Theoretical performance}

To assess the feasibility of SmartphoneDemocracy for large-scale elections, we analyze the computational and network overhead. For TrustChain, we assume a peer-to-peer gossip network. Transactions are broadcast and validated by peers. Furthermore, we assume eventual consistency and validation in accordance with the protocol rules. The primary constraints are the data size stored on the P2P network and the time required for ZKP generation on the smartphone.

We estimate transaction sizes based on standard cryptographic parameter sizes. The BLS12-381 pairing-friendly curve is assumed for BBS.
For the cryptographic primitives, we will use the following estimates:

\begin{itemize}
    \item BBS Proof (\texttt{proof\_reg}): A proof with a few hidden attributes on the BLS12-381 curve is compact but larger than a single zk-SNARK proof. We estimate its size at \({\sim} 1.0\) KB.
    \item ZKP Proof (\(\pi_{\text{vote}}\), Groth16): Remains constant at \({\sim} 200\) bytes.
    \item HE Ciphertext: \({\sim} 512\) bytes.
    \item Commitment/Nullifier: \({\sim} 32\) bytes.
    \item TrustChain Tx Overhead (Signature, public key, etc.): \({\sim} 300\) bytes.
\end{itemize}

This leads to the following approximate transaction sizes per voter:
\begin{itemize}
    \item \(Tx_{\text{reg}}\): \(300 + 32 (\text{cm}) + 1000 (\text{proof}_{\text{reg}}) \approx 1.3\) KB.
    \item \(Tx_{\text{vote}}\): \(300 + 512 (\text{c}) + 32 (\text{nf}_{\text{vote}}) + 200 (\pi_{\text{vote}}) \approx 1.1\) KB.
    \item \(Tx_{\text{share}}\): \(300 + 512 (\sigma) + 200 (\pi_{\text{share}}) \approx 1.0\) KB.
\end{itemize}

The total data generated by an election with \(V\) voters and \(P\) tallying participants is summarized in Table \ref{tab:perfromance_evaluation}.

\begin{table}[ht]
\centering
\caption{Estimated network data size for an election with \(V\) voters, using BBS for registration.}
\label{tab:perfromance_evaluation}
\begin{tblr}{
  width = \linewidth,
  colspec = {lccX},
  row{1} = {c},
  row{6} = {c,m},
  column{even} = {c},
  column{3} = {c},
  vline{2-4} = {-}{},
  hline{2,6} = {-}{},
}
\textbf{Phase}         & \textbf{\# of Txs} & \textbf{Size per Tx} & \textbf{Total Size}                                         \\
\textbf{Config}        & 1                  & ${\sim} 5-10$KB        & ${\sim} 10$ KB (negligible)                                   \\
\textbf{Registration}  & $V$                & ${\sim} 1.8$ KB        & $V \times 1.3$ KB                                           \\
\textbf{Voting}        & $V$                & ${\sim} 1.1$ KB        & $V \times 1.1$ KB                                           \\
\textbf{Tally Share}   & $P$ ($P \ge t$)     & ${\sim} 1.0$ KB        & $P \times 1.0$ KB                                           \\
\textbf{Approx. Total} & $2V + P$      &  N/A                 & $ V \times 2.4 \text{ KB} + P \times 1.0 \text{ KB}$ 
\end{tblr}
\end{table}

For a large-scale election with \textit{1 million voters} (\(V=10^6\)) and 100 tallying participants, the total data stored on the blockchain would be approximately \textbf{2.4 GB}. While this is a substantial figure, it remains well within a manageable range for modern smartphones and P2P network peers, especially considering that nodes may not need to retain the complete data for all past elections indefinitely.

The most intensive task for the smartphone is ZKP proof generation. Based on existing benchmarks for zk-SNARK (Groth16) \cite{oude2024zkpbenchmark}, proving times for circuits of the complexity required for our protocol (hashing, commitment checks, encryption checks) should be well within acceptable limits for a responsive user experience on a modern smartphone. Verification is significantly faster (around \textit{2-10 milliseconds}), ensuring that network peers are not overburdened. 

This analysis shows that the protocol is computationally and network-wise feasible for deployment in real-world, large-scale elections.

\subsection{Evaluation of the performance}

To evaluate the practical performance of the SmartphoneDemocracy protocol, we adopted a micro-benchmarking methodology focused on the system's core cryptographic bottlenecks. Given the complexity of a complete end-to-end implementation, this approach enabled a precise analysis of the most computationally intensive operations that directly impact user experience and may cause bottlenecks in a mobile environment. A Raspberry Pi was selected as the hardware platform for these tests, serving as a representative, resource-constrained device that provides a proxy for the performance one might expect on a modern smartphone.

The experimental setup consisted of a \texttt{Raspberry Pi 5 Model B Rev 1.1} with 8GB of RAM, running Debian 12 for the \texttt{aarch64} architecture. All cryptographic benchmarks were run on Rust version 1.85 and compiled in release mode to ensure optimal performance. To test the BBS signatures used during the registration phase, and Groth16 proofs, we used an implementation of both made available in \cite{githubDockNetworkLibrary}. This setup allows for a realistic and reproducible measurement of the cryptographic primitives at the heart of our protocol.

\begin{figure}
    \centering
    \includegraphics[width=1\linewidth]{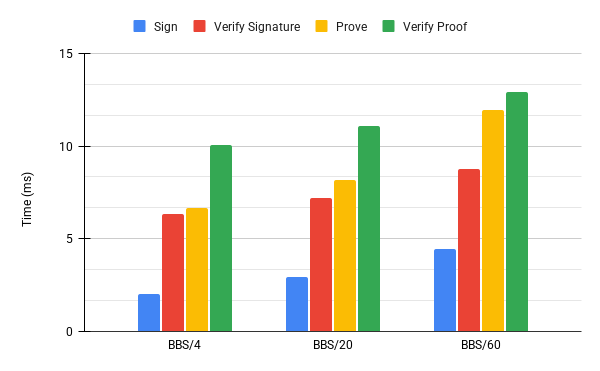}
    \includegraphics[width=1\linewidth]{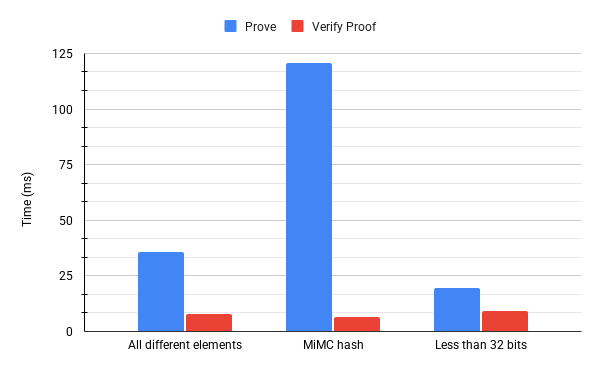}
    \caption{Benchmarks of BBS and ZKP performance performed on Raspberry Pi as a reference for a typical user device.}
    \label{fig:rpi_benchmark}
\end{figure}

Our benchmarks reveal a very satisfactory performance profile from the user's perspective, as seen on Fig. \ref{fig:rpi_benchmark}. On the first graph, we compare the BBS signatures and their performance for different actions, based on the amount of contained credentials, to visualize the scalability of our proposal. It is clearly visible that all actions can be performed in just milliseconds, which is a crucial element of scalability, where each personal device would have to verify proofs of all other registered users. This cost per election is well within acceptable limits for a responsive user application.

\begin{figure}
    \centering
    \includegraphics[width=0.5\linewidth]{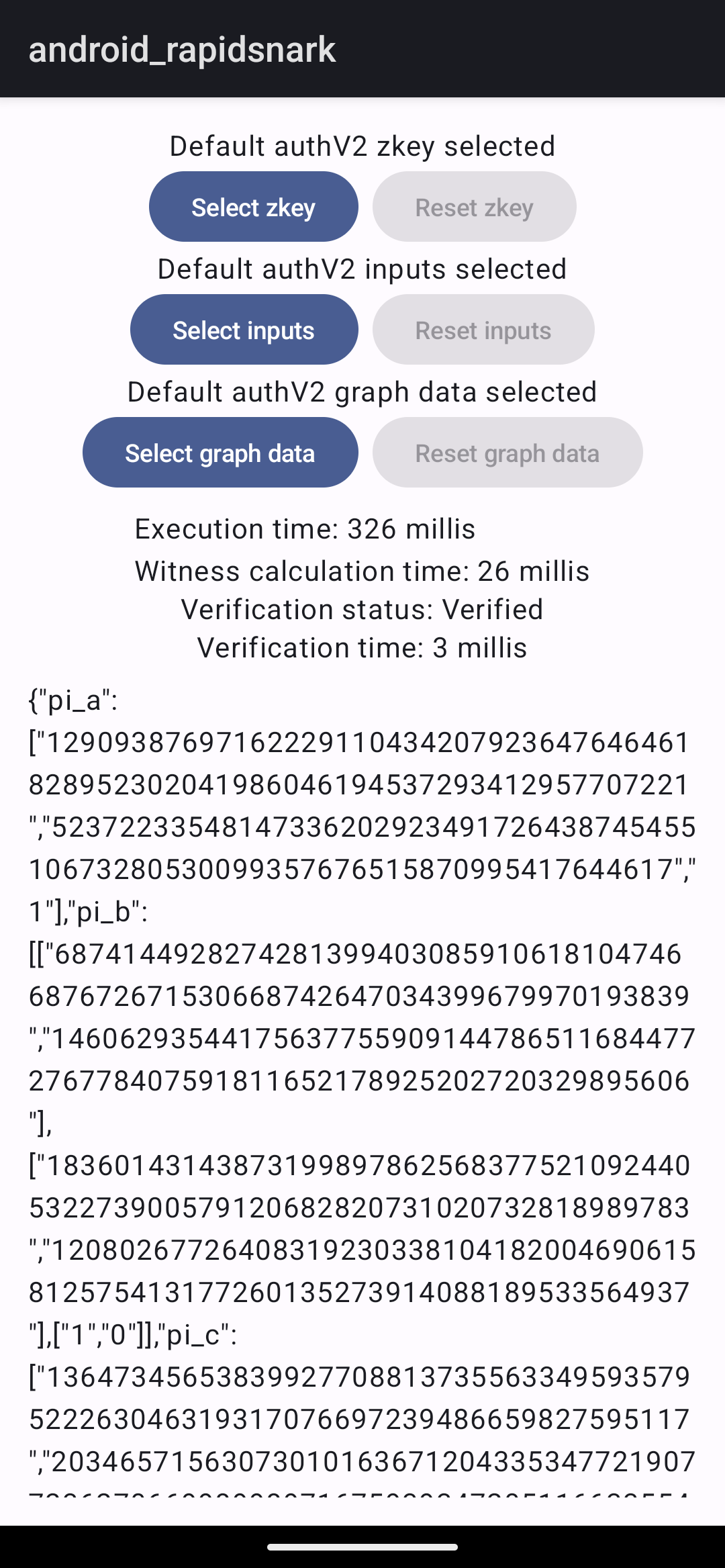}
    \caption{Example app using Android Rapidsnark for proof generation and verification on-device.}
    \label{fig:android_rapidsnark}
\end{figure}

The more computationally intensive task, generating the Groth16 zero-knowledge proof for the vote transaction, required a magnitude higher amount of processing time on our test hardware, but still took under one second for every example. The choice of examples should also be representative of the operations required to verify vote validity. It is also important to note that this is a one-time cost per election, which is again well within acceptable limits, such that the device is not perceived as slow or unresponsive during the critical act of voting. The verification times were consistently measured in the low milliseconds. As an additional confirmation, we also used a direct implementation of experimental Android library called \texttt{Android Rapidsnark}\footnote{Available at: \url{https://github.com/iden3/android-rapidsnark}} to test the actual proving capabilities on an emulated device. Fig. \ref{fig:android_rapidsnark} further confirms our larger-scale testing performed on Raspberry Pi and proves the feasibility.

This high throughput for verification indicates that a single peer device, even one with modest resources like a Raspberry Pi, can validate a significant volume of transactions from other participants. This suggests that the network is unlikely to become bottlenecked by cryptographic verification, supporting the protocol's scalability to a large number of concurrent users. These benchmarks, however, validate the computational feasibility of individual cryptographic operations but do not assess system behavior under concurrent load, network partition scenarios, or the whole end-to-end user experience, including error handling, which is an essential next step.

In summary, the micro-benchmarks of the core cryptographic engines provide strong evidence that the SmartphoneDemocracy protocol is not only theoretically sound but also practically feasible for large-scale deployment.

\subsection{Usability Evaluation}
\label{sub:usability_evaluation}

While a complete usability study is beyond the scope of this initial design, we have assessed the prototype using a personal phone and emulators to get a general overview of the proof of concept. The developed Android prototype application implements the voter-facing components of the protocol. We have ensured that users can view all relevant information on the list of available elections. When clicking on an individual election, they will receive more detailed information and a direct voting interface. Our design prioritizes simplicity, abstracting away the cryptographic complexity into a few button presses, mirroring familiar mobile application workflows (as shown on Fig. \ref{fig:app_usage}).

Based on this prototype, we have conducted most of the general testing of the architecture to ensure it functions as intended from a technical standpoint. All the basic requirements and checks, including the registration flow, voting phase, and final tally, have been implemented, with placeholders in place. This was the crucial element of verifying some of the basic assumptions and a necessary foundation for any future work and improvements. The exact implementation details and a link to the code were provided in Section \ref{sub:implementation_details}.

When considering user experience, we would need to conduct a small field trial with test users to assess their usage and understanding of the system. Preferably, it would include a set list of tasks, a standardized questionnaire such as the System Usability Scale (SUS) to gather feedback, and a brief general interview to understand users' trust in the system. This could take a similar shape to the work done in \cite{waspi2022provotumdesign}, which reflects on Provotum's design (described in Section \ref{sub:provotum}).

\section{Related Work}
\label{sec:related_work}

Literature review on that topic is extensive. Reference \cite{alown2025enhancing} provides a comprehensive survey of electronic voting systems, examining three main approaches: Direct Recording Electronic (DRE), Internet voting, and blockchain-based systems. Their work systematically analyzes the cryptographic foundations of e-voting, including homomorphic encryption, blind signatures, zero-knowledge proofs, and mix-nets, demonstrating how these primitives address core requirements such as voter privacy, verifiability, and coercion resistance. The authors identify blockchain technology as particularly promising for e-voting due to its immutable, transparent, and decentralized architecture. However, they acknowledge challenges in scalability that Layer 2 solutions aim to address. Their comparative analysis reveals that while DRE systems offer convenience and Internet voting provides accessibility, blockchain-based approaches generally excel in providing stronger privacy protections, universal verifiability, and tamper resistance. The paper highlights that an ideal e-voting system must balance multiple competing requirements and suggests that hybrid approaches, leveraging blockchain's security features with other technologies' usability aspects, may represent the most viable path forward for secure, transparent, and widely accepted electronic elections. 

The trends in blockchain-based electronic voting systems have been further studied in \cite{pawlak2021trends}. Through analysis of 52 selected research papers published between 2015 and 2020, the authors identify emerging patterns in how blockchain technology is being applied to address e-voting challenges. Their findings reveal that Ethereum is the most commonly used blockchain platform, followed by Hyperledger Fabric, with many studies not specifying a particular blockchain implementation. The review categorizes e-voting scenarios based on scale (general voting, national voting, small-scale voting, IoT voting) and supervision level (supervised, semi-supervised, unsupervised). Most research focuses on semi-supervised remote voting that can be applied at various scales. The authors note several key cryptographic solutions employed across studies, including digital signatures, zero-knowledge proofs, and homomorphic encryption, which address fundamental e-voting requirements like ballot privacy, eligibility verification, and result verifiability. Despite blockchain's inherent benefits of immutability and decentralization, the review highlights persistent challenges in e-voting implementations, including concerns about coercion resistance, receipt-freeness, transaction fees on public blockchains, and scalability issues. The authors conclude that while blockchain offers promising foundations for e-voting systems, it requires complementary cryptographic techniques and thorough testing to address all the requirements of e-voting. Private permissioned blockchains show particular promise for future implementations.

Both of the aforementioned papers highlight the emerging benefits of blockchain used in the context of e-voting. To analyze it further, we compared some of the existing solutions and highlighted the differences with our approach.
Table \ref{tab:voting_comparison} provides a high-level comparison of our work against several prominent systems, which exposes that the challenge of secure e-voting has been approached from many angles.

\begin{table*}[ht]
\centering
\caption{Comparison of the different voting protocols. REV - Remote Electronic Voting, LEV - Local Electronic Voting (voting station), BB - Bulletin Board}
\label{tab:voting_comparison}
\begin{tblr}{
  width = \linewidth,
  colspec = {lccX[p]X[p]X[p]},
  rowspec = {QQQQQQQ[m]},
  row{1} = {c},
  column{2} = {c},
  column{3} = {c},
  vline{2-6} = {-}{},
  hline{2,7} = {-}{1pt},
}
\textbf{Name}          & \textbf{Blockchain} & \textbf{Vote Type} & \textbf{Core Technology}                                                    & \textbf{Security Properties}                                   & \textbf{Trust Assumptions}                                          \\
\textbf{Provotum \cite{killer2020provotum}}      & Permissioned     & REV                & Smart contracts, DKG, HE                                & Secrecy, transparency, verifiability                  & Authorized block‐signers; no integrated identity layer                \\
\textbf{ElectionBlock \cite{electionblock2021}} & Permissioned        & LEV                & Blockchain ledger, fingerprint auth                                         & Immutability, user control; limited anonymity         & Central operator; permissioned nodes; physical presence                       \\
\textbf{Alethea \cite{alethea2018}}       & None (Perm. BB)     & REV                & Hash‐based codes, NIZK proofs, public randomness                            & E2E verifiability, receipt‐freeness, anonymity        & Honest bulletin board, honest devices                      \\
\textbf{SecureBallot \cite{agate2021secureballot}}  & None                & LEV                & Standard encryption, digital signatures, VPN                            & Secrecy, eligibility, receipt-freeness, verifiability & Honest officials, external notary, secure polling station                          \\
\textbf{zkVoting \cite{cryptoeprint:2024/1003}}      & Public (Ethereum)   & REV                & Homomorphic nullifiable commitments, zk-SNARKs                              & Privacy, coercion resistance, E2E verifiability       & Honest registrar, trusted setup, BBB integrity             \\
\textbf{This paper}    & Public (Trustchain) & REV                & DKG, threshold cryptography, HE, zk-SNARKs, EUDI wallet & Privacy, E2E verifiability, anonymity                 & Honest election setup, Trusted EUDI infrastructure 
\end{tblr}
\end{table*}

\subsection{Provotum}
\label{sub:provotum}

Provotum \cite{killer2020provotum} is a blockchain-based Remote Electronic Voting (REV) system designed to enhance the security, transparency, and verifiability of the voting process. It incorporates several key components, including a public bulletin board that operates on a public permissioned blockchain, where only authorized entities can sign blocks while allowing public verification of all data. The system also utilizes smart contracts, distributed key generation, and homomorphic encryption to ensure ballot secrecy and facilitate verifiability in zero-trust environments. Despite its innovative design, the current implementation of Provotum faces issues due to a lack of an established identity layer and its reliance on a permissioned blockchain, which decreases the decentralization of the system. However, it still provides an excellent framework for the possible design of the system, which could be further modified to suit our needs and applied to our problem description.

\subsection{ElectionBlock}

ElectionBlock \cite{electionblock2021} is an electronic voting system that leverages blockchain technology and fingerprint authentication to enhance voting integrity and security. It addresses the challenges of traditional voting methods, particularly in large-scale elections, by ensuring data immutability and user control over ballots. The system operates on a centralized network of nodes, integrating biometric scanning to distinguish between registered and unregistered voters. While the system analyzes the issues of scalability and performance, it lacks an extensive evaluation of security. The blockchain platform is supposedly designed to handle the properties of anonymity and trust; however, without an explicit solution, the user's anonymity can be revealed by the operator of a permissioned blockchain.

\subsection{Alethea}

Alethea \cite{alethea2018} is a provably secure random sample voting protocol that addresses the unique challenges of polling a small, randomly selected subset of voters. It uses a public bulletin board to publish voter codes, sample group selections derived from a publicly verifiable randomness source (e.g., stock market data), and encrypted ballots, while off-loading critical cryptographic operations such as hash-based code generation and NIZK proofs of plaintext equivalence to personal devices, explicitly modeling human voters, their devices, and compromised platforms as separate roles. It achieves formalized end-to-end verifiability, receipt-freeness, and sample-group anonymity. Its trust assumptions include an honest bulletin board and honest personal devices. In contrast, the voting server is only trusted for privacy (i.e., in threshold-cryptography abstractions) and may be adversarial with respect to integrity. However, using a permissioned bulletin board instead of a decentralized blockchain can make the system more centralized, which means we need clear voter checks to prevent fraud from going unnoticed.

\subsection{SecureBallot}

SecureBallot \cite{agate2021secureballot} is a secure, open-source, supervised e-voting system deployed in polling stations using off-the-shelf laptops, tablets, and PCs connected via a dedicated VPN to a central virtual ballot box. Voter identification and voting phases are fully decoupled with anonymous NFC-based unlocking tokens that activate specific voting booths without ever linking votes to voter identities. Ballots are encrypted on voting stations using AES-CBC with per-ballot symmetric keys and encapsulated under the election’s public key, with HMAC-SHA256 and digital signatures ensuring message integrity and authenticity. All inter-component communications are secured by TLS over IPsec/VPN, and votes are stored atomically via distributed database transactions to guarantee transparency and accuracy. 
Under a strong adversary model, SecureBallot satisfies secrecy, privacy, eligibility, uniqueness, authenticity, integrity, receipt-freeness, and both individual and universal verifiability, formalized through Casper/FDR analyses. Trust assumptions include an honest virtual ballot box, honest polling station staff, and custody of the election’s private key by an external notary until tallying begins. While this centralized design simplifies integration with existing workflows, it forgoes the decentralization of blockchain-based schemes. It introduces single points of trust that demand rigorous checks to bound the risk of undetected fraud. 

\subsection{zkVoting}

ZkVoting \cite{cryptoeprint:2024/1003} is a coercion‐resistant and end‐to‐end verifiable e-voting system built on a novel homomorphic nullifiable commitment scheme and zero-knowledge proofs. It leverages a public blockchain-based bulletin board smart contract (BBB) to publish casting keys, ballots, and proofs, ensuring transparent and tamper-evident logging while preserving voter anonymity. During registration, each voter obtains exactly one real and multiple indistinguishable fake commitment keys from a trusted registrar, enabling them to submit decoy ballots under coercion, which are efficiently nullified at tally time. Ballots are hybrid‐encrypted under the election’s public key and accompanied by Groth16 zk-SNARK proofs to guarantee ballot privacy, integrity, and E2E verifiability. Through formal security proofs, zkVoting satisfies ballot privacy, voter anonymity, receipt‐freeness, coercion resistance, individual and universal verifiability, and eligibility verifiability, with linear tally complexity. The design assumes an honest authority for key issuance, a trusted zk-SNARK trusted setup for public parameters, and the integrity of the blockchain bulletin board smart contract. While offering strong security and practical performance (2.3 seconds per ballot cast, 3.9 milliseconds per ballot tally), its reliance on a centralized registrar introduces single points of trust and scalability trade-offs again. However, its approach to coercion-resistance, while adding considerable complexity, solves the issue quite elegantly.

\section{Discussion}
\label{sec:discussion}

Following the technical design and analysis, this section provides a critical discussion of the SmartphoneDemocracy protocol. It addresses the protocol's current limitations, outlines key avenues for future research, and contextualizes the broader implications of the findings for decentralized democratic systems.

\subsection{Limitations and Future Work}
\label{sub:future_work}

A primary avenue for future work is the complete decentralization of the Verifier role. Our current model relies on a single trusted Verifier that internally manages a nullifier registry, representing a point of centralization. A fully zero-trust architecture could be achieved by replacing this central registry with a public smart contract on a censorship-resistant blockchain. In this model, Verifiers would be stateless entities whose only role is to issue BBS credentials. The voter would then anonymously interact with the smart contract, submitting a BBS proof to register their nullifier on-chain. This approach, while more complex, would eliminate the need to trust any single entity for the registration process. The efficiency and standardized structure of BBS proofs make them exceptionally well-suited for such on-chain verification, minimizing transaction costs and complexity compared to custom ZKP circuits. The governance of such a system, including the management of the authorized verifier list, could itself be decentralized through a Decentralized Autonomous Organization (DAO).

Another crucial element limiting the full assessment of the protocol is the incomplete implementation of all its features. As such, the next step would be to fully develop the necessary underlying cryptographic structure and integrate it with the rest of the app to enable larger-scale experimentation. This is necessary, as most of the analysis in this research is based on microbenchmarks, and assessing real-world performance is crucial to fully verify all our assumptions. The approach we currently use still provides relevant information and demonstrates initial feasibility; however, more specific implementation details may only become apparent during a longer development cycle, which was not possible within the time frame of this research. With this implemented prototype, a real usability evaluation would be performed to gather feedback on the user experience, test the system under load, and gather opinions on the trust in such a system.

For the changes in the general structure, another approach could also involve comparing other blockchain technologies that might scale more effectively and offer additional properties. This research focused on using TrustChain due to its integration within the Super App environment, which allowed for quicker prototyping and satisfied all the basic requirements. There are, however, other options, including smart contract technology, which could be integrated for a multi-layered approach and stronger security properties. This, however, was not possible to properly verify due to time constraints and a lack of experience with these technologies.

Furthermore, there exist more approaches that provide a very convenient solution to the issue of local verification of a significant amount of messages and raw data, such as sharding and delegation. With these ideas, we would be able to distribute some of the workload to other peers, such that our device would not have to process all the ballots. However, with this approach, we would also strengthen our trust assumptions, as we would need to trust the delegated peers to provide us with accurate data. There exist ways to verify certain parts of the delegated calculations, but this would significantly complicate the existing protocol in its current form. This idea could be combined with the multi-chain approach to create layers with smaller distributed trust; however, this is outside the scope of this research.

\subsection{Discussion of Results}
\label{sub:discussion_results}

Our investigation into SmartphoneDemocracy places this work at a unique and exciting moment in technological development. We are witnessing a surge of innovation in foundational fields, such as Zero-Knowledge Proofs and Decentralized Identity. This is not just theoretical; it is accompanied by the rapid emergence of powerful new software libraries that make these advanced concepts accessible to developers. Our work is a direct result of this trend, demonstrating how these cutting-edge tools can be synthesized into a functional system. The experimental stage of many of these libraries signifies a field in dynamic growth, opening up possibilities that were purely academic just a few years ago.

This wave of privacy-preserving software has clearly found its home in the Rust programming language. Our decision to use the cryptographic core in Rust was a strategic one, reflecting its establishment as the de facto standard for high-performance, secure systems development. The language's guarantees of memory safety, combined with an extensive ecosystem of advanced cryptographic libraries, provided the ideal foundation to translate our protocol from theory into a practical prototype. This synergy between cutting-edge research and a robust implementation language is key to the next generation of decentralized applications.

As these powerful technologies become increasingly practical, a significant finding of our work is that the focus must expand beyond purely technical challenges. The successful implementation of a system like SmartphoneDemocracy shifts the conversation towards its social implications. Questions of digital literacy, user interface design, and accessibility become essential. A system that is cryptographically secure but cannot be easily or safely used by the general public has not fulfilled its democratic purpose. This indicates that future work in this domain must treat human-computer interaction and social science as equal partners to cryptography.

This entire endeavor is set against the backdrop of fluctuating public trust in established institutions. This social reality is not an issue, but a feature that our design embraces. Unlike many solutions that aim to digitize existing government processes, SmartphoneDemocracy is engineered for a skeptical world. It is built on the principle of "don't trust, verify." The key implication of our research is that it is now feasible to build systems that derive their legitimacy from public verifiability and cryptographic proof, rather than from faith in an intermediary. It offers a tangible blueprint for a more resilient, direct, and citizen-controlled form of democracy, where participation is a verifiable right, not a granted privilege.

The adoption of SmartphoneDemocracy would face significant regulatory hurdles in most jurisdictions, as electoral laws rarely accommodate novel cryptographic systems without extensive certification and validation. A practical adoption pathway would likely involve smaller-scale applications first (community voting, organizational governance) while building the evidence base and regulatory understanding needed for broader deployment. The emergence of the EUDI Wallet as a legally recognized identity mechanism creates a unique opportunity for this gradual approach to adoption within the European context. 

\section{Conclusion}
\label{sec:conclusion}

In an era of declining institutional trust, this research presented a fundamental re-imagination of e-voting as a permissionless, citizen-centric protocol. We introduced SmartphoneDemocracy, a novel architecture designed to run on a modern smartphone, shifting the core functions of the democratic process away from central authorities. This work provided the protocol's complete end-to-end design, leveraging the European Digital Identity framework, ZKPs, and homomorphic encryption on a peer-to-peer ledger. We formally analyzed its security guarantees, including anonymity and Sybil resistance, and demonstrated through performance evaluation that it is feasible for large-scale deployment on consumer devices.

The core achievement of SmartphoneDemocracy is demonstrating that the mechanics of voting can be securely decoupled from centralized control. Our system serves as a practical blueprint for executing democratic processes within a decentralized web of trust formed directly by citizens. While anchoring initial identity to trusted issuers, we have pushed the boundary of trust significantly and outlined a path toward fully trustless models in future work. This research takes a definitive step toward redefining democracy not as a system granted by institutions, but as a fundamental, self-enforcing right accessible to anyone with a smartphone.

\printbibliography

\appendices

\section{Security Analysis}
\label{app:proofs}
Table \ref{tab:risk_analysis} provides an overview of the risk analysis for the SmartphoneDemocracy protocol.

\newpage
\begin{table*}
\centering
\caption{Risk analysis for the E-voting implementation. P - probability, I - impact}
\label{tab:risk_analysis}
\begin{tblr}{
  width = \linewidth,
  colspec = {Q[c,m]XXQ[c,m]Q[c,m]},
  row{1} = {c,m},
  column{-} = {m},
  cell{2}{1} = {r=3}{},
  cell{5}{1} = {r=5}{},
  cell{10}{1} = {r=4}{},
  cell{14}{1} = {r=4}{},
  cell{18}{1} = {r=3}{},
  cell{21}{1} = {r=3}{},
  vline{2-5} = {1-24}{},
  hline{2,5,10,14,18,21} = {-}{},
  hline{3-23} = {2-5}{},
}
\textbf{Threat Category}     & \textbf{Threat}                                                               & \textbf{Mitigation}                                                                                          & \textbf{P} & \textbf{I} \\
\textbf{Eligibility Reg.}    & \textbf{1a. Compromised EUDI Issuer (issuing false eligibility VCs)}          & Reliance on EU standards trusted issuer lists; Verifier cross-checks.                                       & Low        & High       \\
                             & \textbf{1b. Compromised/DoS Verifier Server (blocking/allowing invalid regs)} & Server security, redundancy; ZKP/Nullifier check on TrustChain provides a secondary barrier.                 & Medium     & Medium     \\
                             & \textbf{1c. Stolen EUDI credentials used for reg.}                            & EUDI Wallet security (biometrics/PIN); Verifier interaction requires live session; Nullifier prevents reuse. & Low        & Low        \\
\textbf{Vote Casting}        & \textbf{2a. Coercion (forcing vote choice or token handover)}                 & Potential for deniability/re-voting. Focus on ZKP hiding choice.                                             & Medium     & Medium     \\
                             & \textbf{2b. Malware on phone (stealing secrets, changing vote pre-ZKP)}       & Secure OS, app sandboxing, user awareness.                                                                   & Medium     & High       \\
                             & \textbf{2c. ZKP flaw allowing invalid vote proof (wrong format/range)}        & Formal analysis of ZKP vote circuit.                                                                         & Low        & High       \\
                             & \textbf{2d. TrustChain Censorship/DoS (preventing vote submission)}           & P2P network redundancy, multiple peer connections.                                                           & Medium     & Medium     \\
                             & \textbf{2e. Usability issues preventing vote casting}                         & Good UI/UX design, testing, help guides.                                                                     & Medium     & Low        \\
\textbf{Vote Secrecy}        & \textbf{3a. HE scheme broken / flawed implementation}                         & Use of established, well-analyzed HE schemes (Paillier etc.) and implementations.                            & Low        & High       \\
                             & \textbf{3b. Collusion of HE key share holders (=t participants)}              & Threshold cryptography (t required), decentralized holding of shares, ZKP proof of correct share usage.      & Low        & High       \\
                             & \textbf{3c. Side-channel attack on phone revealing vote/secrets}              & OS-level security, constant-time crypto implementations (where possible).                                    & Low        & Medium     \\
                             & \textbf{3d. Network analysis linking voter identity to TX}                    & Use of mixnets/Tor (adds complexity), P2P diffusion.                                                         & Medium     & Medium     \\
\textbf{Vote Counting / Tally} & \textbf{4a. HE flaw leading to incorrect sum calculation}                     & Mathematical properties of HE scheme.                                                                      & Low        & High       \\
                             & \textbf{4b. Insufficient tally participants (t) provide shares}               & Incentives for participation, setting reasonable t, public monitoring of participation rate.                 & Medium     & High       \\
                             & \textbf{4c. Malicious tally participants submit invalid shares}               & ZKP proof is required for each share, verified by anyone combining shares.                                      & Low        & Medium     \\
                             & \textbf{4d. ZKP flaw allowing bad shares}                  & Formal analysis of ZKP circuit. Redundancy due to the threshold.                                                                    & Low        & High       \\
\textbf{System Integrity}    & \textbf{5a. TrustChain network failure/partition}                             & P2P resilience, but a large-scale internet outage is possible.                                                 & Low        & High       \\
                             & \textbf{5b. EUDI Wallet/Verifier infrastructure failure}                      & Redundancy, standard protocols allow potential alternative verifiers.                                        & Low        & Medium     \\
                             & \textbf{5c. Large-scale smartphone malware/OS vulnerability}                  & Ecosystem diversity helps, but a major 0-day is conceivable.                                                   & Low        & High       \\
\textbf{Transparency / Audit}  & \textbf{6a. Complexity hides potential flaws (ZKP/HE opaque to public)}       & Publish code/circuits, formal verification, explanations. Relies on expert trust.                          & High       & Medium     \\
                             & \textbf{6b. TrustChain data inaccessible/unverifiable}                        & Public nature of blockchain (if readable), independent verification tools.                                   & Low        & High       \\
                             & \textbf{6c. Trusted setup compromise for zk-SNARKs}                           & Use transparent setup schemes (MPC-based), or alternatives like zk-STARKs (trade-offs).                      & Low        & High       
\end{tblr}
\end{table*}

\end{document}